# Stochastic Models of Quantum Mechanics – A Perspective

Mark P. Davidson

*Spectel Research Corp. 807 Rorke Way, Palo Alto, CA 94303, USA,*
*Email: mdavid@spectelresearch.com*

**Abstract.** A subjective survey of stochastic models of quantum mechanics is given along with a discussion of some key radiative processes, the clues they offer, and the difficulties they pose for this program. An electromagnetic basis for deriving quantum mechanics is advocated, and various possibilities are considered. It is argued that only non-local or non-causal theories are likely to be a successful basis for such a derivation.

## INTRODUCTION

The motivation for stochastic models is the belief that quantum theory is incomplete. The essential goal is summed up well even today by Einstein "*There is no doubt that quantum mechanics has seized hold of a good deal of truth, and that it will be a touchstone for any future theoretical basis ... However, I do not believe that quantum mechanics can serve as a starting point in the search for this basis, just as vice versa, one could not find from thermodynamics the foundations of mechanics*" [1]. The program is nothing less than a derivation of quantum mechanics from something like classical physics, probably based on some combination of point particles, solitons, classical strings, classical field theory, and using techniques from statistical mechanics, complexity theory, chaos theory, and nonlinear theory. Stochastic models are hidden variable theories. As evidence of increasing interest, figure 1 shows citations of David Bohm's landmark 1952 papers. In my opinion there is no complete stochastic derivation or explanation of quantum theory at the present time, and a suitable candidate theory that might be the complete answer is not known. An interesting step in this direction has recently been proposed by Adler et al. [2], which shows how quantum theory may evolve as a thermodynamic limit of a "classical system" of non-commuting matrices. The importance of this work is that it reduces the gap between classical theories and quantum theory, and it is sufficiently general as to embrace many different types of stochastic theories, provided they can be cast in the form of classical non-commuting matrices. There are a number of tantalizing bits and pieces to the puzzle which await a master stroke to put them together into a coherent whole. In this paper I hope to present a mosaic of clues to a stochastic derivation of quantum mechanics, and to point out some of the many difficulties facing this effort.

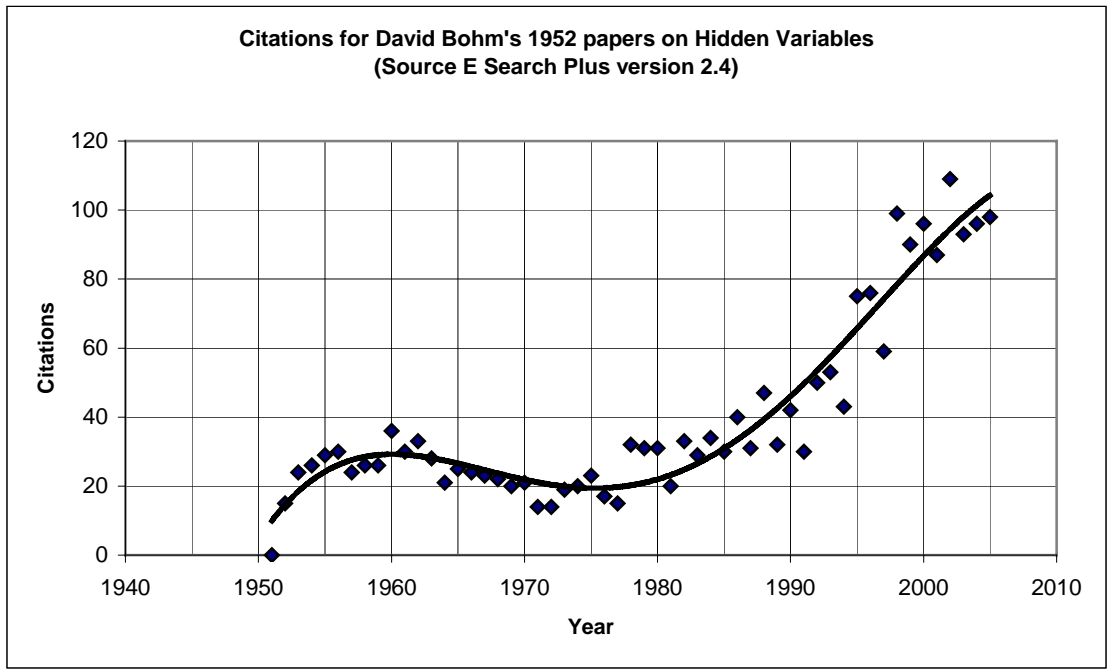

**FIGURE 1.** A measure of activity in hidden variable and stochastic research

It would be wrong to give the impression that there is not still strong opposition to hidden variables within the physics community, including a stochastic program for quantum mechanics. For example, R. F. Streater on his website [3] places hidden variable theories under the heading "Lost Causes in Theoretical Physics" and writes about them "Bell's theorem, together with the experiments of Aspect et al., shows that the theoretical idea to use hidden classical variables to replace quantum theory is certainly a lost cause, and has been for forty years." It is clear from this conference alone that Streater's statement would not be supported by many of the participants even though most of them do not subscribe to a hidden-variable viewpoint. However, his opinion still mirrors the views of most theoretical physicists today. Despite this, throughout the history of quantum theory the hidden variable viewpoint simply refuses to die out and far from dying out it is gaining prominent adherents such as Gerard 't Hooft [4-7], Stephen Adler [2], and Lee Smolin [8,9] in recent years. The quantum interpretation problem has impacted the foundations of probability theory as well [10].

# CAN ELECTROMAGNETISM BE THE ORIGIN OF QUANTUM MECHANICS?

It's an old idea that the stochastic origins of quantum mechanics may be electromagnetic, and by this I mean based somehow on classical electromagnetism. This is a minority opinion even among advocates of ontological quantum models, but it is surprising how many unusual features of electromagnetism are coming to light in recent years, how much mystery there is still to be found in classical electromagnetism, and how many quantum phenomena have an electromagnetic interpretation. The first and perhaps most compelling piece evidence supporting an electromagnetic origin is the fact that the fine structure constant is a dimensionless number

$$\alpha = e^2/\hbar c \text{ (in cgs units)} = 7.297\ 352\ 568(24) \times 10\text{-}3 \qquad (1)$$

And as a consequence, Planck's constant can be expressed in terms of purely electromagnetic constants. Nelson stressed this fact in [11] (page 65).

Early proposals for an electromagnetic interpretation for quantum mechanics include the semiclassical and neoclassical radiaton theories of Jaynes [12-15]. A detailed review of these and other related topics can be found in the excellent text by de la Pena and Cetto [16]. The successes and limitations of semiclassical radiation theory were pointed out in [17]. Another version of the semiclassical theory with some modifications is due to Carver Mead [18], illustrating how practical engineers, needing quantum theory in their work, balk at the philosophical baggage which they have been handed by physicists. Other work along these lines are the early stochastic electrodynamic papers which treat the vacuum fluctuations of quantum electrodynamics in the ground state as a real classical random electromagnetic field [16] and which evolved into stochastic electrodynamics which we discuss below.

## Quantum Wave equations as generators of classical non-radiating electromagnetic sources

Non-radiating electromagnetic sources have a long history in physics with practical applications in plasma physics, optics, electrical engineering, inverse scattering, and with fundamental importance to a number of physical phenomena. The earliest commonly cited references are [19-23]. The most general form for such current distributions is not known. There has been an effort to describe elementary particles as extended objects of this type [21-26]. And assorted other references may be found in [27].

In several recent papers [27,28] it has been shown with some generality that the currents generated by free-particle wave equations are non-radiating sources. The results are proven for non-relativistic as well as for relativistic wave equations. We reproduce here the results for Schrödinger's equation only and refer the reader to [27,28] for more details.

**Definition** A non-radiating source is a 4-current $J^\mu(x)$ for which the integrated Poynting vector over a spherical surface $S_R$ of radius $R$ vanishes in the limit of large radius for all time using only the retarded fields due to it.

$$\lim_{R\to\infty} \frac{4\pi}{c} \iint_{S_R} \mathbf{E}_{ret}(x, t+R/c) \times \mathbf{B}_{ret}(x, t+R/c) \cdot \mathbf{dS} = 0 \text{ for all t.} \qquad (2)$$

Consider the free-particle Schrödinger equation and its charge and current densities.

$$-\frac{\hbar^2}{2m}\Delta\Psi = i\hbar\frac{\partial\Psi}{\partial t} \tag{3}$$

$$\rho(\mathbf{x},t) = q\Psi^*\Psi; \quad \mathbf{J}(\mathbf{x},t) = \frac{q\hbar}{2mi}\left\{\Psi^*\nabla\Psi - \Psi\nabla\Psi^*\right\} = \frac{q\hbar}{2mi}\left\{\Psi^* \overset{\leftrightarrow}{\nabla} \Psi\right\} \tag{4}$$

$$\tilde{\Psi}(\mathbf{p},t) = \int \Psi(\mathbf{x},t)e^{-i\mathbf{p}\cdot\mathbf{x}/\hbar}d^3x, \quad \Psi(\mathbf{x},t) = \frac{1}{(2\pi\hbar)^3}\int \tilde{\Psi}(\mathbf{p},t)e^{i\mathbf{p}\cdot\mathbf{x}/\hbar}d^3p \tag{5}$$

**Theorem 1**

Let the wave function $\tilde{\Psi}(\mathbf{p},0)$ at some initial time be in the Schwartz class $S(R^3)$ so that it is everywhere infinitely differentiable and falls off faster than any power, and moreover let it have compact support such that

$$\tilde{\Psi}(\mathbf{p},0) = 0, \text{ for } |\mathbf{p}| > p_{max} = (1-\varepsilon)mc \text{ for some } \varepsilon > 0, \tilde{\Psi}(\mathbf{p},0) \in S(R^3) \tag{6}$$

This condition insures that both the phase velocity $|\mathbf{p}|/2m$ and the group velocity $|\mathbf{p}|/m$ are less than the speed of light. Then the current generated from the Schrödinger wave is a non-radiating source. The proof is given in [27,28].

The Klein-Gordon equation must be restricted to either positive energy or negative energy solutions in order to get non-radiating sources. The Dirac equation is more tricky. One must first make a Foldy-Wouthuysen transformation in order to get a non-radiating source [27,29]. For higher-spin equations, if they can be cast in the canonical form of Foldy [27,30], then non-radiating source currents are naturally generated by them. Quantum field theorists may not find these results surprising, as there are many similarities between the classical radiation theory and the quantum radiation theory, and of course quantum free particles do not radiate in quantum electrodynamics (QED). The quantum mechanical and classical calculations aren't completely equivalent because the photon energy is quantized in the quantum theory, but it is not quantized in the classical theory. Thus there is an energy threshold that must be overcome for emission of quantum radiation of a given frequency, but there is no such threshold classically. This may explain why the Dirac current actually leads to radiation classically (unless one first performs a Foldy-Wouthuysen transformation), but does not in QED.

There is debate in the literature over the acceptability of relativistic wave equations. Prugovecki [31] has argued that acceptable relativistic local wave equations cannot exist, and he has proposed a phase-space alternative which is based on a smearing function called a resolution generator. The transformation to phase space looks something like a short time Fourier transform. Prugovecki's phase-space wave equations generate many more conserved currents which depend on the resolution generator, and it can be shown that these are all non-radiating sources too for free particles [27].

## A stochastic interpretation of the zero radiation result

Consider a random classical radiation field. The most famous example of such a field is the zero-point radiation of stochastic electrodynamics, but I don't want to limit this discussion to it. Let a classical charged particle move with response to this field and possibly to other unspecified random forces. Assume that the equation of motion is such that the incoming field averages to zero and the presence of the particle does not produce any net radiation on the average.

$F_{vacuum}^{\mu\upsilon}(x,t)$ is a random electromagnetic field satisfying

$$E(F_{vacuum}^{\mu\upsilon}(x,t)) = 0 \qquad (7)$$

Now imagine placing a charged particle in this vacuum which will move under the force and will also modify the vacuum fields

$$F^{\mu\upsilon}(x,t) = F_{vacuum}^{\mu\upsilon}(x,t) + F_{rad}^{\mu\upsilon}(x,t) \qquad (8)$$

$$F_{rad}^{\mu\upsilon}(x,t) = \partial^{\mu}A_{rad}^{\upsilon} - \partial^{\upsilon}A_{rad}^{\mu}; \quad A_{rad}^{\mu}(x,t) = \frac{1}{c}\int\frac{J^{\mu}(\mathbf{x'},t-\frac{R}{c})}{R}d^{3}x'; \quad R = |\mathbf{x}-\mathbf{x'}| \qquad (9)$$

If the particle is in equilibrium with the vacuum, it is necessary that it doesn't radiate on the average. Equilibrium will of course depend on the details of the radiation field, and the equation of motion of the particle, but without specifying these, we can still learn something. If we take an ensemble average, the incoming wave averages to zero and it follows that.

$$E(F_{rad}^{\mu\upsilon}(x,t)) = o\left(\frac{1}{R}\right) \text{ and so therefore } E(J^{\mu}(\mathbf{x'},t-\frac{R}{c})) \text{ is non-radiating} \qquad (10)$$

This is just what we have found for Schrödinger's free-particle equation. We thus have a qualitative explanation for why Schrödinger's equation should have this non-radiating property. The radiationless behavior is a natural relaxation mechanism that can be understood in a stochastic model. In the Copenhagen interpretation is appears as simply a coincidence. A similar argument can be made for neutral particles that will generally have some non-vanishing magnetic or electric moments of higher order. The free-particle quantum wave equations are non-radiating for these objects as well if the electromagnetic fields are treated classically and the moments decouple from the particle's motion.

## Some Low energy bremsstrahlung results

The results in Table 1 for the hydrodynamic and QED case were derived in [32], the other cases are straightforward. For the quantum electrodynamic calculation the quantized electromagnetic field was used. For the other cases classical fields were used in the calculations of bremsstrahlung. The acceleration **a** in these formulae is just the applied external force divided by mass. David Bohm's quantum mechanical force is given by

$$F_{QM} = -\nabla \left[ -\frac{\hbar^2}{2m} \frac{\Delta\sqrt{\rho}}{\sqrt{\rho}} \right] \qquad (11)$$

| Classical Radiation Result | $E_{rad} = \frac{2}{3}\frac{q^2}{c^3} \int_0^T \mathbf{a}^2(t')dt'$ |
|---|---|
| Hydrodynamic Model Result | $E_{rad} = \frac{2}{3}\frac{q^2}{c^3} \int_0^T \left|\langle\Psi|\mathbf{a}(t')|\Psi\rangle\right|^2 dt'$ |
| Quantum electrodynamics | $E_{rad} = \frac{2q^2}{3c^3} \int_0^T \langle\Psi|\mathbf{a}^2(t')|\Psi\rangle dt'$ |
| Bohmian mechanics | $E_{rad} = \frac{2}{3}\frac{q^2}{c^3} \int_0^T \int \left(\mathbf{a}(t') + F_{QM}/m\right)^2 \rho(x)d^3x dt'$ |
| Newtonian ensemble | $E_{rad} = \frac{2}{3}\frac{q^2}{c^3} \int_0^T \int \mathbf{a}(t')^2 \rho(x,p,t') d^3x d^3p dt'$ |
| Stochastic Mechanics | $E_{rad} = \infty$ |

**Table 1** Low energy bremsstrahlung formulae showing dependence on the wave function

A hydrodynamic model disagrees with quantum electrodynamics for low energy bremsstrahlung. In [32] experimental tests were proposed which could discriminate between the two. QED acts like a point particle as far as bremsstrahlung is concerned. Feynman put it this way: "The wave function for an electron in an atom does not, then, describe a smeared-out electron with a smooth charge density. The electron is either here, or there, or somewhere else, but wherever it is, it is a point charge" [33]. We see also that a naive application of classical electromagnetism gives the wrong answer for bremsstrahlung for Bohmian mechanics and stochastic mechanics (although Inclusion of a fluctuating vacuum field could allow the stochastic mechanics result to be simply counted as rearranging the vacuum field). Theoreticians interested in constructing a stochastic model of quantum mechanics should keep these results in mind.

## A derivation of Schrödinger's equation from the Lorentz-Dirac equation coupled with Gibbs' distribution

The following argument is the only one that the author knows which yields Schrödinger's equation within the context of classical statistical mechanics [34]. Consider a charged particle described by Lorentz-Dirac equation in the nonrelativistic limit [35].

$$m_0 \mathbf{a}(t) = \int_0^\infty ds\, e^{-s} \mathbf{F}(\mathbf{x}(t+\tau s), t), \quad \tau = 2q^2/3m_0 c^3 \tag{12}$$

Suppose that the vacuum is alive with random field fluctuations, and suppose that it has a small temperature $T$. A more precise definition of this concept will not be attempted. It will only be assumed that the classical Gibbs distribution is satisfied. If the radiative force were ignored, then the particle would reach a state of equilibrium at temperature $T$, and its spatial density would be given by the classical Gibbs distribution,

$$\rho(x) = e^{-V(x)/kT}, \quad kT \nabla \ln(\rho) = -\nabla V = \mathbf{F}_{\text{ext}} \tag{13}$$

These equations would not be satisfied by a charged particle experiencing a significant radiative force. The statistical distribution in this case is simply not known. Two assumptions shall be made to include radiative forces in the simplest possible way. The charged particle, in thermal equilibrium with the vacuum, is described by a continuous Markov process, the same process as is used in stochastic mechanics and Brownian motion [34,36,37]. Using Nelson's notation, $\mathbf{x}$ is assumed to satisfy the stationary stochastic differential equation

$$d\mathbf{x} = \mathbf{b}(\mathbf{x}(t))dt + d\mathbf{W}(t) \tag{14}$$

where $W$ is a three-dimensional Wiener process with

$$E(dW_i(t) dW_j(t)) = 2\nu \delta_{i,j} dt \tag{15}$$

Consider the following conditional expectation:

$$\mathbf{F}_E(\mathbf{x}) = -E\left( \int_0^\infty ds\, e^{-s} \nabla V(\mathbf{x}(t+\tau s)) \Big| \mathbf{x}(t) = \mathbf{x} \right) \tag{16}$$

This expresses the expected value of the total force on the particle, including pre-acceleration, given that at time $t$ the particle's trajectory passed through the point $\mathbf{x}$. It represents the best estimate that can be made of the instantaneous force acting on the particle at position $\mathbf{x}$ and time $t$. By analogy with the classical Gibbs distribution, the following equation for the charged particle is postulated:

$$kT \nabla \ln(\rho) = \mathbf{F}_E(\mathbf{x}) \tag{17}$$

The Markov transition function expressed as a density satisfies the "forward" and "backward" equations of Kolmogorov:

$$P_{t-u}(\mathbf{y},\mathbf{x}) = \lim_{d^3y \to 0} P(\mathbf{x}(t) \in d^3\mathbf{y} | \mathbf{x}(u) = \mathbf{x}) / d^3y, \text{ where } \mathbf{y} \in d^3y$$

$$\frac{\partial}{\partial t}P_{t-u}(\mathbf{y},\mathbf{x}) + \nabla_y \cdot (\mathbf{b}(\mathbf{y})P_{t-u}(\mathbf{y},\mathbf{x})) - \upsilon\Delta_y P_{t-u}(\mathbf{y},\mathbf{x}) = 0, \quad t > u, \text{ forward eq.} \quad (18)$$

$$\frac{\partial}{\partial u}P_{t-u}(\mathbf{y},\mathbf{x}) + \mathbf{b}(\mathbf{x}) \cdot \nabla_x P_{t-u}(\mathbf{y},\mathbf{x}) + \upsilon\Delta_x P_{t-u}(\mathbf{y},\mathbf{x}) = 0, \quad t < u, \text{ backward eq.}$$

$\mathbf{F_E}$ may be expressed in terms of the Markov transition function, and so

$$\mathbf{F_E}(\mathbf{x}) = -\int_0^\infty ds\, e^{-s} \int d^3y \nabla V(\mathbf{y}) P_{\tau s}(\mathbf{y},\mathbf{x}) = kT \ln(\rho(\mathbf{x})) \quad (19)$$

This expresses the pre-acceleration integral in terms of the Markov transition function. These relations lead to the following Schrödinger-like equation [34] after a somewhat intricate derivation

$$[-2\tau\upsilon kT\Delta + V + 2kTR]e^R = \lambda e^R, \quad \lambda = \text{a constant}, R = \ln(\rho)/2 \quad (20)$$

This can be rewritten as

$$\rho(\mathbf{x}) = \exp\left[-\left(V(\mathbf{x}) - 2\tau\upsilon kT(\Delta\sqrt{\rho(\mathbf{x})})/\sqrt{\rho(\mathbf{x})}\right)/kT\right] \quad (21)$$

And the effective force can be written as

$$\mathbf{F_E}(\mathbf{x}) = -\nabla\left[V(\mathbf{x}) - 2\tau\upsilon kT(\Delta\sqrt{\rho(\mathbf{x})})/\sqrt{\rho(\mathbf{x})}\right] \quad (22)$$

If it happens that

$$\frac{\hbar^2}{2m} = 2\tau\upsilon kT = \frac{4}{3}\frac{q^2\upsilon kT}{m_0 c^3} \quad (23)$$

Then we get Schrödinger's equation, but with an extra nonlinear term

$$[-(\hbar^2/2m)\Delta + V + 2kTR]e^R = \lambda e^R \quad (24)$$

In [34] it was suggested that T be identified with the cosmological microwave background temperature which today is estimated at about 2.725 K°. Unfortunately, there is no experimental evidence for a nonlinear term in Schrödinger's equation. Neutron diffraction experiments [38,39] set the limit for the nonlinear term at

$$|kT| < 3.3 \times 10^{-15} eV \quad or \quad |T| < 3.829 \times 10^{-11} K^o \quad (25)$$

To give a sense of scale, this small energy is equal to the electrostatic energy of two electrons separated by a distance of 435 km. Nonlinear wave equations were first proposed in [40]. It can be shown that the presence of the log term does not lead to any bremsstrahlung radiation for a free particle to lowest order in the multipole expansion. But the higher order terms will radiate in general. And so perhaps the reason why the log term is absent has to do with the need to have zero radiation to all orders in the multipole expansion.

Wallstrom [41] has pointed out that stochastic mechanics is ambiguous in that it does not necessarily explain angular momentum conservation and single valuedness of the wave function. In the current model, which is consistent with generalized stochastic mechanics [37] the wave function in (24) must be purely real and since it must also be continuous it follows that for a spherically symmetric potential that angular momentum will be quantized if the nonlinear term is ignored, thus offering a resolution of Wallstrom's objection to stochastic mechanics in the present context.

## Other properties of the Lorentz-Dirac Equation

There is a great deal that is unknown about the Lorentz-Dirac equation [35] – the leading candidate for the equation of motion of a point charge particle. Phenomena possibly predicted by it include: Planck radiation formula derived if zero-point radiation present [42,43]; Both reflection and transmission at potential barriers with the same initial conditions [44]; Non-uniqueness of solutions [44,45]; Bell type nonlocality [46]; An analog of particle pair creation and annihilation [47]; Tunneling phenomena [47]; Aharanov-Bohm effect [48]; Interesting Hydrogen-like solutions [49].

## Some very brief comments on stochastic electrodynamics

Stochastic electrodynamics is a vast topic, I refer the reader to the best review on the subject [16]. It gives an appealing qualitative picture of quantum mechanics, explaining quantum indeterminism. It works reasonably well for linear systems like charged particle in a harmonic oscillator well and predicts the ground state wave function convincingly in these cases. It describes the Casimir force nicely. It provides a mechanism for particle inertia as due to interaction between charged particle and the zero point field. A possible explanation of gravity as an emergent phenomena has been proposed which is interesting and closely related to an electromagnetic model of gravity proposed by Dicke some years ago [50-52]. Stochastic electrodynamics is appealing conceptually, and it has attracted a number of sincere and ardent adherents.

Problems exist though [16]. Isolated particles or systems in quantum mechanics have their energy distribution strictly conserved. If they are in an energy eigenstate they remain so. But a particle in SED gains or loses energy due to interaction with the zero point field. Atoms tend to spontaneously ionize in SED as a consequence. The Lorentz-Dirac equation is too intractable to allow a clean and precise derivation of Schrödinger's equation with the kind of precision required by the neutron diffraction tests for general nonlinear potentials or even for free particles as far as I am aware. The spectral absorption and emission lines are too broad in simple calculations published so far to come anywhere close to fitting the myriad of atomic spectral data.

A clean and rigorous derivation of the Planck radiation formula for general nonlinear mechanical systems is still lacking to my knowledge. Van Vleck's exhaustive work [53,54] still poses a problem for this endeavor. It is surprising that light coming from very distant reaches of the universe hasn't lost all coherent effects if it had to pass through billions of light years of classical zero point radiation. Any causal electromagnetic model has difficulty explaining the elementary radiative process where one atom emits a single photon and another distant atom absorbs this same photon. In classical radiation theory, this type of behavior is inconceivable. Something is clearly missing if we are to derive quantum mechanics from electromagnetism.

## Wheeler-Feynman electrodynamics

One of the things to try is the action at a distance theory pioneered by Wheeler and Feynman [55-57]. Maxwell's equations are invariant under time reversal, but whenever we integrate them to calculate the fields generated by a current, we must use the retarded potentials instead of the advanced ones in order to agree with reality. We never hear a radio show before it is broadcast, only afterwards. Nature is causal, but Maxwell's equations are not necessarily causal unless we impose causality on them by choosing only the retarded solutions. Wheeler and Feynman modified electromagnetism by averaging over advanced and retarded effects with equal weight. The resulting theory is non-causal, and as such is capable of resolving all problems with hidden variables and Bell's theorem. For example, Cramer's transactional interpretation is based on a generalization of Wheeler-Feynman theory [58], and suggests that a fully electromagnetic model based on Wheeler-Feynman might possibly explain quantum entanglement and the various nonlocality paradoxes including radiation and absorption of radiation by well-separated atoms.

## Tachyon stochastic models are a possibility

Although considered somewhat radical, tachyons interacting electromagnetically are interesting [59-61]. They could explain quantum non-locality trivially. They can explain quantum tunneling, and it seems that they can form bound states that might be very hard to break. A very interesting paper by Waite, Barut, and Zeni has shown that purely electromagnetic particles can exist as solitons if they are made up of tachyonic fluids [25]. They also gives a proof of an interesting theorem for such solutions that field lines will move slower than light even though currents are moving faster. This could explain why signals cannot be sent faster than light even though superluminal connections seem to be present in quantum mechanics. This theory also results in radiationless motion of the currents, and so the solutions are always non-radiating sources. David Bohm and Jean-Pierre Vigier were both advocates of real superluminal motions as a basis for quantum mechanics [62,63]. Other forms of nonlocality are discussed by Namsrai [64].

# STOCHASTIC MECHANICS AND TRACE DYNAMICS

Trace Dynamics, formally similar to classical mechanics, deals with non-commuting matrices as the basic objects [2]. Canonical quantization rules may emerge independent of the form of the non-commutivity of the basic matrix elements, as they follow from statistical mechanical arguments which are similar to the equipartition theorem [2]. I believe trace dynamics can be related to and can complement Nelson's stochastic mechanics [11,36,65]. The position of a classical Brownian particle is simply a c number vector variable, and the positions commute at different times. However there is a temporal non-commutative structure to Brownian motion and it's generalizations that is due to the fact that the sample trajectories of the Wiener process are not differentiable. Consider the Wiener process two-point correlation function.

$$E(w(t_1)w(t_2)) = 2\nu \min(t_1, t_2) \tag{26}$$

This last equation is an indication that there is something strange about this process. Taking a derivative with respect to $t_1$ for example we find

$$\frac{\partial}{\partial t_1} E(w(t_1)w(t_2)) = \begin{Bmatrix} 2\nu \text{ if } t_1 < t_2 \\ 0 \text{ if } t_1 > t_2 \end{Bmatrix} \tag{27}$$

If we take the limit where $t_1$ approaches $t_2$ from either direction we find two different results. It is a classical result that the time derivative of the Wiener process does not exist [66]. In fact the trajectories of the Wiener process and of stochastic mechanics are known to be Fractal curves [67] with dimension 2. This result shows a temporal non-commutivity that can be the basis of a reformulation of the stochastic process in terms of non-commuting operators similar to the Heisenberg representation in quantum mechanics [68]. It provides a way to link the trace dynamics theories to stochastic mechanics. Trace dynamics needs non-commuting matrices. But non-commuting operators on a Hilbert space can be expressed as matrices by choosing an orthonormal basis, and therefore the mathematics of Brownian motion can be recast in terms of non-commuting matrices. An initial attempt at this was made in [65]. So in this way the machinery of trace dynamics can be joined to the machinery of Brownian motion to enhance the applicability of both theories. The benefit of this synergy remains to be explored however.

## The Differential Space Hidden Variable model

It is possible to construct a measurable random trajectory space from quantum mechanics by using the elegant and powerful theory of Wiener and Siegel [69-74]. It is also possible to generalize this theory to allow for many random trajectory measures all of which are experimentally indistinguishable from one another, but which provide different stochastic trajectories for each observable of a quantum mechanical system [65]. This generalization allows one to construct many stochastic models of quantum mechanics, all of which predict the correct quantum mechanical outcomes of arbitrary measurements. The original Wiener-Siegel theory was deterministic, but non-deterministic stochastic models can be constructed as well [65]. It is likely that stochastic mechanics can be included within this framework. The arena of the theory is the same Hilbert space as quantum theory. Besides the usual state vector $\Psi$ which pertains to an ensemble of similarly prepared states, there is another vector in the Hilbert space, call it $\alpha$, which is associated with each element of the ensemble and which is treated like a random variable. Unlike $\Psi$ which is fixed, $\alpha$ is a random variable to be averaged over. The probability space for $\alpha$ is called differential space. One advantage that the Wiener-Siegel theory has over stochastic mechanics is that it provides a stochastic process for every single observable operator, whereas stochastic mechanics singles out the coordinate variable.

## Conclusion

The effort to derive quantum mechanics by stochastic means spans many years and methods. The first task is to find a stochastic explanation for the quantum laws as they are currently understood, and then to go beyond them and predict something new. The kinetic theory of gases had a similar history. At first it just reproduced the ideal gas laws along with other known thermodynamic equations. Eventually it predicted new phenomena such as in Einstein's Brownian motion theory. The problem is that a stochastic derivation of quantum theory has proven to be much more difficult than the kinetic theory of gases. The conceptual obstacles are profound and confounding. Turning away from the challenge, and embracing the non-realistic Copenhagen type of interpretations is therefore a temptation. But for some of us the possibility of a realistic derivation of quantum mechanics is simply too compelling to abandon and so we persist despite the difficulties. Some find the results so far disappointing. But to me the glass is half full, not half empty. The large and growing number of suggestive connections between classical stochastic models and an assortment of quantum phenomena especially in classical electromagnetism are simply too powerful a set of clues to ignore. If a derivation of quantum mechanics can be found, we can be certain that it will be one of the greatest triumphs of the human intellect.

## Acknowledgements

The author thanks Andrei Khrennikov, Guillaume Adenier, and Växjö University for their kind hospitality during this conference.